\documentclass{article}
\usepackage{spconf}

\usepackage{graphicx}
\usepackage{booktabs}
\usepackage{amsmath}
\usepackage[table,xcdraw]{xcolor}

\title{AE-Flow: AutoEncoder Normalizing Flow}
\name{Jakub Mosiński, Piotr Biliński, Thomas Merritt, Abdelhamid Ezzerg, Daniel Korzekwa}
\address{Amazon Alexa AI \\ \{mosinjak, bilipiot, ezzerg\}@amazon.com}

\begin{document}
\ninept
\maketitle

\begin{abstract}
Recently normalizing flows have been gaining traction in text-to-speech (TTS) and voice conversion (VC) due to their state-of-the-art (SOTA) performance. Normalizing flows are unsupervised generative models. In this paper, we introduce supervision to the training process of normalizing flows, without the need for parallel data. We call this training paradigm AutoEncoder Normalizing Flow (AE-Flow). It adds a reconstruction loss forcing the model to use information from the conditioning to reconstruct an audio sample. Our goal is to understand the impact of each component and find the right combination of the negative log-likelihood (NLL) and the reconstruction loss in training normalizing flows with coupling blocks. For that reason we will compare flow-based mapping model trained with: (i) NLL loss, (ii) NLL and reconstruction losses, as well as (iii) reconstruction loss only. Additionally, we compare our model with SOTA VC baseline. The models are evaluated in terms of naturalness, speaker similarity, intelligibility in many-to-many and many-to-any VC settings. The results show that the proposed training paradigm systematically improves speaker similarity and naturalness when compared to regular training methods of normalizing flows. Furthermore, we show that our method improves speaker similarity and intelligibility over the state-of-the-art.

\end{abstract}

\section{Introduction} \label{sec:intro}
Voice conversion is the task of transforming speech from a source voice to 
sound as though it was spoken by the desired target voice~\cite{mohammadi2017overview, sisman2020overview}. In other words, we want to change the speaker identity in speech while preserving linguistic information. There are two main data paradigms in voice conversion: the use of parallel~\cite{abe1988vq, shikano1991codebook} and non-parallel training data~\cite{lorenzo2018voice, mouch2006nonparallel, serra2019blow, Merritt2022}. The former assumes access to parallel training data, \textit{i.e.} recordings that differ only in speaker identity. Such data allows the mapping between speakers to be learned with supervision. Unfortunately, real parallel data does not exist. We could use signal processing techniques such as dynamic time warping~\cite{helander2008align} to match the recordings on the frame level, but the quality of such transformed recordings is questionable. The latter paradigm does not require parallel data and utilizes unsupervised learners such as normalizing flows.


In this paper, we investigate normalizing flows following their recent success in text-to-speech (TTS) ~\cite{bilinski2022creating, miao2020flow, kim2020glow, valle2020flowtron, casanova2021sc} and voice conversion (VC)~\cite{serra2019blow, Merritt2022}. Flow-based generative models learn mapping from the input data to a latent vector~\cite{kobyzev2020normalizing, kingma2018glow,  ho2019flowpp}. This mapping is done through a sequence of invertible transformations using the change of variables rule to obtain a valid probability distribution allowing for exact sampling and density evaluation. They explicitly maximize the likelihood of the prior distribution resulting in a stable convergence. Additional conditioning can be provided to the flow steps via coupling blocks to help maximize the likelihood or to achieve additional control over the signal generation~\cite{serra2019blow, miao2020flow, valle2020flowtron, casanova2021sc, ho2019flowpp}. Coupling blocks are constructed in such a way that the information present in the conditioning should be removed when encoding to the latent space and added when decoding from the latent space. That mechanism allows to control the speech features of interest by changing the conditioning information between encoding and decoding procedures. However, the above is true only if the model learns the contribution of the conditioning to the speech sample.

Unfortunately, flow-based generative models do not perfectly disentangle speaker identity from the audio sample, which may come from an inability to fully utilize speaker embedding conditioning~\cite{serra2019blow}. We observe potential speaker information leakage to the latent space by training a speaker classifier on the average pooled latent space across time dimension. A two-layer perceptron classifier achieves 29\% accuracy on a 118-speaker test set indicating that the conditioning might not be fully utilised.

In this work, we propose a new training paradigm of normalizing flows that adds supervision to enforce the use of conditioning and improves speaker similarity between the target speaker recording and audio signal generated by the VC model. The proposed paradigm is called AutoEncoder Normalizing Flow (AE-Flow), which is a normalizing flow VC model trained as an autoencoder with an additional reconstruction loss, \textit{e.g.} L1 loss.
During speech generation, we decode from the sampled prior distribution assuming that all necessary information is provided via coupling blocks. This mitigates source speaker leakage and speeds up inference as we can omit the encoding step.
We hypothesise that this approach enforces the use of conditioning.

We apply the proposed training paradigm to FlowVC model~\cite{bilinski2022creating, miao2020flow} that has demonstrated the state-of-the-art quality.
We study the balance between NLL and L1 reconstruction losses in training normalizing flows.
Moreover, we compare our model with state-of-the-art voice conversion model.
The methods are evaluated in terms of naturalness, speaker similarity, intelligibility in many-to-many and many-to-any VC settings.
The experiments show that the proposed training paradigm systematically improves speaker similarity and naturalness when compared to regular training methods of normalizing flows. Furthermore, we show that our method improves speaker similarity and intelligibility over the state-of-the-art.

\section{Method} \label{sec:method}
\begin{figure*}[h]
\centering
\includegraphics[width=\textwidth]{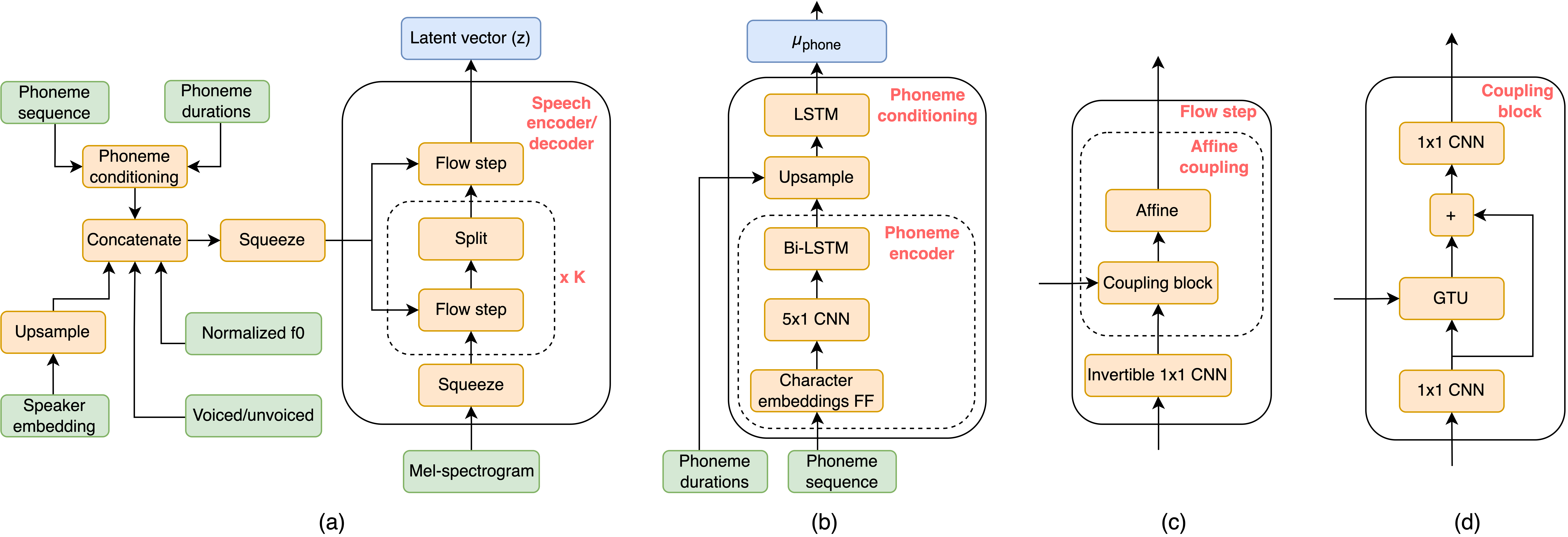}
\caption{Overview of the VC approach using normalizing flows (a) followed by phoneme conditioning component (b),  Flow step (c) and Coupling block (d). The arrows correspond to the direction of mel-spectrogram encoding to the latent space.}
\label{fig:flowttsarchitecture}
\end{figure*}

\subsection{Normalizing Flows}
Normalizing flows aim to approximate an unknown true data distribution $p(x)$ from a set of observations $\{x\}_{i=1}^N$. The flow-based generative model learns a bijective transformation $f_{\theta}(.)$ (where $\theta$ denotes neural network's parameters) that maps a latent space with tractable distribution $p_{\iota}(z)$ to $x$:

\begin{equation}
    z \sim p_{\iota}(z), \; x=f_{\theta}(z) \wedge \, z=f^{-1}_{\theta}(x).
    \label{eq:flows}
\end{equation}
Here we assume that $p_{\iota}$ is a standard normal distribution $\mathcal{N}(0, I)$. What is more,  the normalizing flow $f_{\theta}$ is composed of a sequence of $K$ invertible transformations $f^{-1}_{\theta}=f_{\theta_{1}}^{-1} \circ f_{\theta_{2}}^{-1} \circ \dots \circ f_{\theta_{K}}^{-1}$. One of the major properties of normalizing flows is that they can directly model the density from Equation \ref{eq:flows} under the change of variable theorem. We can compute the exact log-likelihood for a given data point $x$ as:
\begin{equation}
    \log p_{\theta}(x) = \log p_{\iota}(z) + \sum_{i=1}^K\log \Big|\det \frac{\partial  f_{\theta_{i}}^{-1}(x_i)}{\partial x_i}\Big| ,
    \label{eq:loglikelihood}
\end{equation}
where $\frac{\partial  f_{\theta_{i}}^{-1}(x_i)}{\partial x_i}$ is the Jacobian of $f_i^{-1}(x_i)$. Given that we can directly compute $\log p_{\theta}(x)$, the normalizing flow is optimized via negative log-likelihood. This makes the training more stable compared to optimizing adversarial loss in Generative Adversarial Networks (GANs) or the lower bound of the NLL for Variational Autoencoders (VAEs)~\cite{kobyzev2020normalizing}. We can also exactly compute the log-likelihood of a given sample $x$.

To better maximize the log-likelihood and obtain control over speaker identity we use conditional normalizing flows and closely follow the architecture of FlowVC~\cite{bilinski2022creating, miao2020flow}, shown in Figure \ref{fig:flowttsarchitecture}. Conditioning is provided via a coupling layer. The encoding to $z$ looks as follows:
\begin{equation}
    \psi = \{ ph_{source}, vuv_{source}, f0_{source} \},
    \label{eq:additionalconditioning}
\end{equation}
\begin{equation}
    z = f_{\theta}^{-1}(x; spk_{source}, \psi),
    \label{eq:conditionalencodeing}
\end{equation}
where $spk_{source}$ is a pre-trained  mean  speaker embedding corresponding to the source speaker~\cite{wan2018generalized}, $ph_{source}$ is a phoneme conditioning  coming  from  the phoneme encoder. We extract phoneme sequence and durations as in~\cite{shah2021}. Following, $vuv_{source}$ is a binary value denoting whether a frame is voiced or unvoiced, $f0_{source}$ is a sentence-level mean normalised interpolated log-f0. The f0 normalization is applied to remove speaker identity (\textit{i.e.},  relating to the speaker’s average f0)  from  sentence  prosody,  thus  separating  f0  conditioning  from speaker embedding conditioning.

Finally, to perform voice conversion we first sample the latent space $z$ from a prior distribution, see Equation \ref{eq:vcsampling}. Then, to generate mel-spectrogram in a target voice $x_{gen}$, we use mean speaker embedding of the target speaker $spk_{target}$ and other features extracted from the source speech, see Equation \ref{eq:vcconditionaldecoding}:
\begin{equation}
    z \sim p_{\iota}(z),
    \label{eq:vcsampling}
\end{equation}
\begin{equation}
    x_{gen} = f_{\theta}(z; spk_{target}, \psi).
    \label{eq:vcconditionaldecoding}
\end{equation}

\subsection{AutoEncoder Normalizing Flow}
Normalizing flows have many useful properties such as exact log-likelihood estimation, stable convergence and meaningful latent representation. Passing speaker embedding conditioning though the coupling blocks allows to learn how to add or remove speaker information to the stream of flows.
Unfortunately, the flow-based generative models do not perfectly disentangle speaker identity from the audio sample, see Section~\ref{sec:intro}.
We hypothesise that by adding supervision we could strengthen the signal from the conditioning and improve speaker similarity of a flow-based mapping model. 

In this section, we introduce AutoEncoder Normalizing Flow, a new paradigm for training normalizing flows with additional losses. This work focuses on L1 reconstruction loss, but the approach could be generalized to other losses such as L2 or adversarial loss. AE-Flow first encodes a mel-spectrogram $x$ to the latent space $z$, see Equation \ref{eq:conditionalencodeing2}. Then, the $z'$ is sampled from a normal distribution, see Equation \ref{eq:sampling}, and used for the decoding to the mel-spectrogram to obtain $x_{gen}$, see Equation \ref{eq:conditionaldecoding2}.
\begin{equation}
    z = f_{\theta}^{-1}(x; spk_{source}, \psi),
    \label{eq:conditionalencodeing2}
\end{equation}
\begin{equation}
    z' \sim p_{\iota}(z),
    \label{eq:sampling}
\end{equation}
\begin{equation}
    x_{gen} = f_{\theta}(z'; spk_{source}, \psi).
    \label{eq:conditionaldecoding2}
\end{equation}
Notice that the conditioning for the encoding and decoding is exactly the same. The sampling step is necessary, otherwise, if we would not exchange $z$ to $z'$ for decoding then $x_{gen}=x$ as $f_{\theta}(f^{-1}_{\theta}(x))=x$ and the additional loss would be meaningless. Finally, we can write the objective function of the AE-Flow:
\begin{equation}
   \frac{1}{N}\sum_{i=1}^{N}\big [ -(1-\lambda)\log p_{\theta}(x^{(i)}) + \lambda \|x_{gen}^{(i)} - x^{(i)}\|_1 \big ],
    \label{eq:aeloss}
\end{equation}
where $\lambda$ is a hyperparameter that controls the balance between losses.

\section{Experimental Setup}
\subsection{Dataset}
We use Amazon's internal high-quality dataset. The US English professional voice talents were asked to read the provided text in a recording studio. The training set has 118 gender-balanced speakers. There are approximately 91k utterances with an average recording length of 3.9s. A sampling rate of 24 kHz was used for all recordings, from which 80-dimensional mel-spectrograms were extracted using a frame shift of 12.5 ms.  For generating audio samples for evaluation, the Universal Neural Vocoder was used~\cite{jiao2021universal}. To evaluate the systems we use two datasets: 
\begin{itemize}
    \item S to S (seen source speaker to seen target speaker) - 5 male and 5 female speakers randomly chosen from the training data. For each source speaker we take 20 different utterances not seen during training. Then we create all source speaker, target speaker, utterance combinations and randomly choose 600 of them for evaluation.
    \item S to U (seen source speaker to unseen target speaker) - we use the same source speakers as in the S to S dataset. There are 4 male and 4 female target speakers unseen during training. Finally, we create a dense conversion mapping from seen to unseen speaker and randomly choose 600 combinations for evaluation. 
\end{itemize} 

\subsection{Evaluated Systems}
To study the proposed training paradigm, we compare the following models: AE-Flow, FlowVC, ND-Flow and CopyCat~\cite{karlapati2020CopyCat}. 
The AE-Flow uses both NLL and L1 reconstruction losses. To select the reconstruction loss weight parameter $\lambda$ (see Equation \ref{eq:aeloss}), we considered $\lambda \in \{0.5, 0.9, 0.99\}$ and chose the best performing $\lambda=0.99$ based on the internal subjective preference tests. 
FlowVC, a normalizing-flow model trained only with NLL is used as a baseline approach ($\lambda=0$ in Equation \ref{eq:aeloss}). 
We introduce ND-Flow, a noise decoding flow that uses the same architecture as AE-Flow and FlowVC, but is trained from the Gaussian noise to the target mel-spectrogam utilizing only L1 loss ($\lambda=1$ in Equation \ref{eq:aeloss}). We trained our models up to 100 epochs with batch size 64 and Adam optimizer~\cite{kingma2014adam} on two Tesla V100 16GB GPUs with PyTorch 1.10.2+cu102~\cite{torch2019} and frozen random seed. Finally, we include CopyCat model as a state-of-the-art non flow-based VC baseline. Throughout the work we will refer to source speaker recordings as Source and non-parallel target speaker recordings as Target.

\subsection{Evaluation Protocol} \label{Evaluation Protocol}
To measure performance and compare voice conversion models we use the following metrics:
\begin{itemize}
    \item Speaker similarity: MUSHRA evaluation~\cite{series2014method}, where people are given the following instruction: ``Please listen to the speaker in the reference sample first. Then rate how similar the speakers in each system sound compared to the reference speaker''. Two different recordings from the target speaker are included. One as the reference sample and the other as one of the systems to be rated, as an upper-anchor. In addition, the source speech recording is included among the systems to be rated, as a lower-anchor. 
    \item Naturalness: MUSHRA evaluation where people are given the following instruction: ``Please  rate  the  audio  samples  in  terms  of their naturalnes''. The recording from the source speaker is included among the systems to be rated as an upper-anchor.
    \item Word Error Rate (WER): To measure the intelligibility of converted speech we conduct WER analysis.  It was computed by comparing the original text of an utterance with the transcription obtained by the ASR system of the converted speech using a pre-trained kaldi TDNN chain model~\cite{kaldi2011}.
\end{itemize}
For each MUSHRA evaluation there were 240 testers with 20 ratings per tester. Significant differences between systems were detected using paired t-tests with Holm-Bonferroni correction applied. All reported significant differences are for $p_{value}~\leq~0.05$.

\section{Experimental Results}
\subsection{Comparison of Flow-based Approaches}

\begin{table}
\centering
\begin{tabular}{|lcc|}
\hline
\rowcolor[HTML]{C0C0C0} 
\multicolumn{3}{|c|}{\cellcolor[HTML]{C0C0C0}Speaker Similarity}                                                   \\ \hline
\rowcolor[HTML]{EFEFEF}
\multicolumn{1}{|c|}{\cellcolor[HTML]{EFEFEF}System} & \multicolumn{1}{c|}{\cellcolor[HTML]{EFEFEF}S to S} & S to U \\ \hline
\multicolumn{1}{|l|}{FlowVC} & \multicolumn{1}{c|}{$73.8 \pm 0.6$} & $73.4 \pm 0.6$  \\ \hline
\multicolumn{1}{|l|}{AE-Flow} & \multicolumn{1}{c|}{$74.3 \pm 0.6$} & $73.9 \pm 0.6$  \\ \hline
\multicolumn{1}{|l|}{ND-Flow} & \multicolumn{1}{c|}{$74.4 \pm 0.6$} & $73.8 \pm 0.6$  \\ \hline
\multicolumn{1}{|l|}{Source} & \multicolumn{1}{c|}{$70.2 \pm 0.8$} & $71.0 \pm 0.7$  \\ \hline
\multicolumn{1}{|l|}{Target} & \multicolumn{1}{c|}{$75.2 \pm 0.6$} & $74.6 \pm 0.6$  \\ \hline
\rowcolor[HTML]{C0C0C0} 
\multicolumn{3}{|c|}{\cellcolor[HTML]{C0C0C0}Naturalness}                                                          \\ \hline
\rowcolor[HTML]{EFEFEF} 
\multicolumn{1}{|c|}{\cellcolor[HTML]{EFEFEF}System} & \multicolumn{1}{c|}{\cellcolor[HTML]{EFEFEF}S to S} & S to U \\ \hline
\multicolumn{1}{|l|}{FlowVC} & \multicolumn{1}{c|}{$72.7 \pm 0.7$} & $73.3 \pm 0.7$  \\ \hline
\multicolumn{1}{|l|}{AE-Flow} & \multicolumn{1}{c|}{$73.0 \pm 0.7$} & $73.7 \pm 0.7$  \\ \hline
\multicolumn{1}{|l|}{ND-Flow} & \multicolumn{1}{c|}{$73.0 \pm 0.7$} & $73.8 \pm 0.7$  \\ \hline
\multicolumn{1}{|l|}{Source} & \multicolumn{1}{c|}{$73.2 \pm 0.7$} & $74.1 \pm 0.7$  \\ \hline
\end{tabular}
\caption{Average scores and 95\% confidence intervals for speaker similarity and naturalness evaluations comparing FlowVC, AE-Flow and ND-Flow. S to S - seen source speaker to unseen target speaker, S to U - seen source speaker to unseen target speaker.}
\label{tab:mushraflows}
\end{table}

In this experiment we compare three normalizing flow models: FlowVC{ }--{ }no reconstruction loss, AE-Flow{ }--{ }both NLL and L1 reconstruction loss,  ND-Flow{ }--{ }only L1 reconstruction loss to assess the influence of the NLL and L1 reconstruction losses balance on the speaker similarity and naturalness of the generated speech. Evaluation results are presented in Table \ref{tab:mushraflows}. 

Considering the speaker similarity, AE-Flow and ND-Flow improve upon FlowVC, both in the the seen to seen and seen to unseen cases.  It shows that adding L1 loss to the NLL objective improves speaker similarity. There is no statistical significance between AE-Flow and ND-Flow in both cases. During informal listening, we observed that the L1 objective regularizes and prevents extreme behaviour. As an example, FlowVC occasionally makes the generated sample sound too high pitched when converting from female to female voice, thus diverging from the target speaker. This behaviour is less noticeable in the AE-Flow and ND-Flow. We hypothesise the reason for that is the ``averaging'' nature of the L1 loss preventing extreme changes in pitch occasionally occurring in samples generated by FlowVC.

Regarding the Naturalness, models with additional L1 reconstruction loss outperform the model trained only with NLL loss. In the S to S case, only the comparison of FlowVC and Source is statistically significant. The power of the test was too low to reject the hypothesis that both AE-Flow and ND-Flow generated samples are on par with real recordings in terms of the naturalness. In the many-to-any case, both AE-Flow and ND-Flow statistically improve upon FlowVC.

This study shows that adding the L1 objective improves speaker similarity and naturalness when training normalizing flow models. The lack of statistical difference between the AE-Flow and ND-Flow prevents us from conclusively comparing those models. It is wroth mentioning that ND-Flow training is almost two times faster than AE-Flow, as we only need to perform the decoding step, see Equation \ref{eq:conditionaldecoding2}.

\subsection{Comparison to SOTA VC Approach}
\begin{table}
\centering
\begin{tabular}{|lcc|}
\hline
\rowcolor[HTML]{C0C0C0} 
\multicolumn{3}{|c|}{\cellcolor[HTML]{C0C0C0}Speaker Similarity}                                                   \\ \hline
\rowcolor[HTML]{EFEFEF}
\multicolumn{1}{|c|}{\cellcolor[HTML]{EFEFEF}System} & \multicolumn{1}{c|}{\cellcolor[HTML]{EFEFEF}S to S} & S to U \\ \hline
\multicolumn{1}{|l|}{FlowVC} & \multicolumn{1}{c|}{$74.2 \pm 0.7$} & $74.2 \pm 0.7$  \\ \hline
\multicolumn{1}{|l|}{AE-Flow} & \multicolumn{1}{c|}{$74.5 \pm 0.7$} & $74.1 \pm 0.7$  \\ \hline
\multicolumn{1}{|l|}{CopyCat} & \multicolumn{1}{c|}{$73.9 \pm 0.7$} & $73.8 \pm 0.7$  \\ \hline
\multicolumn{1}{|l|}{Source} & \multicolumn{1}{c|}{$71.0 \pm 0.8$} & $71.9 \pm 0.7$  \\ \hline
\multicolumn{1}{|l|}{Target} & \multicolumn{1}{c|}{$75.3 \pm 0.7$} & $75.2 \pm 0.6$  \\ \hline
\rowcolor[HTML]{C0C0C0} 
\multicolumn{3}{|c|}{\cellcolor[HTML]{C0C0C0}Naturalness}                                                          \\ \hline
\rowcolor[HTML]{EFEFEF} 
\multicolumn{1}{|c|}{\cellcolor[HTML]{EFEFEF}System} & \multicolumn{1}{c|}{\cellcolor[HTML]{EFEFEF}S to S} & S to U \\ \hline
\multicolumn{1}{|l|}{FlowVC} & \multicolumn{1}{c|}{$73.5 \pm 0.7$} & $72.5 \pm 0.7$  \\ \hline
\multicolumn{1}{|l|}{AE-Flow} & \multicolumn{1}{c|}{$73.8 \pm 0.7$} & $73.0 \pm 0.7$  \\ \hline
\multicolumn{1}{|l|}{CopyCat} & \multicolumn{1}{c|}{$73.4 \pm 0.7$} & $73.4 \pm 0.7$  \\ \hline
\multicolumn{1}{|l|}{Source} & \multicolumn{1}{c|}{$74.1 \pm 0.6$} & $73.6 \pm 0.7$  \\ \hline
\end{tabular}
\caption{Average scores and 95\% confidence intervals for speaker similarity and naturalness evaluations comparing FlowVC, AE-Flow, CopyCat. S to S - seen source speaker to unseen target speaker, S to U - seen source speaker to unseen target speaker.}
\label{tab:mushrasota}
\end{table}

Further MUSHRA evaluations were conducted to compare the flow-based generative models with the non flow-based SOTA baseline. The MUSHRA scores are used to assess speaker similarity and naturalness of the generated samples in the S to S and S to U scenarios, see Section \ref{Evaluation Protocol}. The evaluation results are presented in Table \ref{tab:mushrasota}. 

Considering speaker similarity, flow-based VC models scored higher that the CopyCat baseline. This shows that flow-based models are on par or superior to the CopyCat model in terms of speaker similarity, as also found in~\cite{bilinski2022creating}. However, statistical significance was achieved only between AE-Flow and CopyCat in the S to S case. This shows that our AE-Flow method improves upon the SOTA non flow-based voice conversion model in terms of speaker similarity.

The naturalness results show no statistical significance between the flow-based models and CopyCat in the S to S case. It is worth mentioning that the MUSHRA evaluators could not distinguish between the AE-Flow and real recordings. In the zero-shot experiment the CopyCat model outperforms the FlowVC, but the comparison to the AE-Flow is not statistically significant.

The experiment shows that the flow-based models can achieve similar performance in voice conversion to non flow-based SOTA methods. The results also suggest that our method outperforms the CopyCat in speaker similarity and naturalness in some settings.

\subsection{Word Error Rate Analysis}
\begin{table}
\centering
\begin{tabular}{|lcc|}
\hline
\rowcolor[HTML]{C0C0C0} 
\multicolumn{3}{|c|}{\cellcolor[HTML]{C0C0C0}Word Error Rate (\%)}                                                   \\ \hline
\rowcolor[HTML]{EFEFEF}
\multicolumn{1}{|c|}{\cellcolor[HTML]{EFEFEF}System} & \multicolumn{1}{c|}{\cellcolor[HTML]{EFEFEF}S to S} & S to U \\ \hline
\multicolumn{1}{|l|}{FlowVC} & \multicolumn{1}{c|}{$12.2 \pm 1.6$} & $11.3 \pm 1.6$  \\ \hline
\multicolumn{1}{|l|}{AE-Flow} & \multicolumn{1}{c|}{$12.4 \pm 1.5$} & $12.8 \pm 1.5$  \\ \hline
\multicolumn{1}{|l|}{ND-Flow} & \multicolumn{1}{c|}{$12.6 \pm 1.5$} & $12.0 \pm 1.4$  \\ \hline
\multicolumn{1}{|l|}{CopyCat} & \multicolumn{1}{c|}{$16.2 \pm 1.7$} & $15.8 \pm 1.6$  \\ \hline
\multicolumn{1}{|l|}{Source} & \multicolumn{1}{c|}{$12.0 \pm 1.6$} & $11.7 \pm 1.6$  \\ \hline
\end{tabular}
\caption{Average word error rates and 95\% confidence intervals for word error rate comparing FlowVC, AE-Flow, ND-Flow, CopyCat. S to S - seen source speaker to unseen target speaker, S to U - seen source speaker to unseen target speaker.}
\label{tab:wer}
\end{table}

In this section, we study intelligibility of generated samples in both S to S and S to U cases. In Table \ref{tab:wer} we gather word error rate scores for the FlowVC, AE-Flow, ND-Flow, CopyCat, and a reference Source recordings.

All flow-based models outperform CopyCat with statistical significance. However, there is no statistically significant difference between FlowVC, AE-Flow and ND-Flow. This may suggest that the architecture has the most significant impact on intelligibility, and not the training loss.

\section{Discussion}
Our experiments show that the addition of reconstruction loss improves upon the standard NLL training of normalizing flows. However, the question arises why not use L1 loss only, as there is no statistical difference between AE-Flow and ND-Flow.
The dataset used in this paper was created by professional voice actors with recordings of similar style. The low variance in a speaker conditioned distribution may mitigate the perception of the ``averaging'' effect of the L1 loss. It is possible that training on larger range of different recording conditions with some speakers recorded in studio-quality conditions whilst other speakers recorded in more ambient surroundings using lower quality microphones, the advantage of AE-Flow over Flow-VC and ND-Flow could become more noticeable. We leave that topic for future work, but suggest to optimise the balance between NLL and L1 losses for a given dataset.

Another direction for future work is considering the use of L2, adversarial or other losses in the AE-Flow training setup. The proposed method is general and not constrained to the L1 loss.

\section{Conclusions}
In this paper we have proposed a new training paradigm of flow-based generative models called AutoEncoder Normalizing Flows that introduces supervision to the training procedure without the need for parallel data. We have comprehensively evaluated our methods and baselines in terms of speaker similarity, naturalness and intelligibility in many-to-many and many-to-any voice conversion scenarios. The results show that adding the L1 reconstruction loss to the normalizing flow training objective improves both speaker similarity and naturalness of the generated samples. Our method also improves upon the non flow-based SOTA CopyCat model in terms of intelligibility and speaker similarity. Moreover, our training method can be easily generalized to other supervised objectives such as L2 loss and adversarial loss.

\newpage
\bibliographystyle{IEEEtran}

\bibliography{mybib}

\end{document}